\documentstyle[12pt]{article}
\begin{document}

\title{The entanglement concentration and purification for entangled polarization photon state}
\author{XuBo Zou, K. Pahlke and W. Mathis  \\
\\Institute TET, University of Hannover,\\
Appelstr. 9A, 30167 Hannover, Germany }
\date{}

\maketitle

\begin{abstract}
{\normalsize We present a concentration and purification scheme
for nonmaximally entangled pure  and mixed porization entangled
state. We firstly show that two distant parties Alice and Bob
first start with two shared but less entangled photon pure states
to produce a four photon GHZ state, and then perform a $45°$
polarization measurement onto one of the two photons at each
location such that the remaining two photon are projected onto a
maximally entangled state. We further show the scheme also can be
used to purify a class of mixed polarization-entangled state.}

\end{abstract}
The entanglement is an essential resource for quantum
communication. Highly entangled state play a key role in an
efficient realization of quantum information processing including
quantum teleportation\cite{chg}, cryptography\cite{ake}, dense
coding\cite{chs}, and computation\cite{chd}. In these protocols,
the maximally entangled state are required. Since the entangled
state prepared for quantum information processing is open to an
environment, the quality of entanglement is easily degraded. To
get maximalliy entangled state from nonmaximally entangle pure and
mixed entangled state, the entanglement concentration and
purification have been proposed\cite{con}. In these protocols, a
smaller number of maximally entangled state can be obtained from
less entangled state using local operations and classical
communication (LOCC). Recently several experimental schemes have
been proposed for concentrating the nonmaximally entangled photon
state\cite{pan,aa}. A more powerful purification scheme working
for general mixed polarization-entangled states has also been
proposed, which is based on the detection of fourfold
coincidence\cite{pa}. Recently, we proposed a scheme to generate
the polarization-entangled state and entangled N photon
state\cite{zou}. In this paper, We firstly show that such scheme
can be used to produce a four photon GHZ state from two shared but
less entangled photon pure states using local operations and
classical communication (LOCC). Two distant parties Alice and Bob
then perform a $45°$ polarization measurement onto one of the two
photons at each location such that the remaining two photon are
projected onto a maximally entangled state. We further show the
scheme also can be used to purify a kind of mixed
polarization-entangled state.\\
Consider the experiment shown schematically in Fig.1, which
produce a four photon GHZ state from two shared but less entangled
photon pure states using local operations and classical
communication (LOCC). Let us assume that Alice and Bob are given
two pairs of photons in the following polarization entangled
state
$$
|\Phi>_{12}=\alpha|H_1>|H_2>+\beta|V_1>|V_2> \eqno{(1)}
$$
$$
|\Phi>_{34}=\alpha|H_3>|H_4>+\beta|V_3>|V_4> \eqno{(2)}
$$
where $|\alpha|^2+|\beta|^2=1$. Alice holds photons 1 and 3, and
Bob holds photon 2 and 4. The total state of the system can be
written in the form
$$
|\Psi>=|\Phi>_{12}|\Phi>_{34}=\alpha^2|H_1>|H_3>|H_2>|H_4>+\beta^2|V_1>|V_3>|V_2>|V_4>
$$
$$
+\alpha\beta(|H_1>|V_3>|H_2>|V_4>+|V_1>|H_3>|V_2>|H_4>) \eqno{(3)}
$$
We forward photon 1 and 3, 2 and 4 to polarizing beam splitter
$PBS_1$ and $PBS_2$, respectively. Since the PBS transmits only
the horizonal polarization component and reflect the vertical
component, after photons 1 and 3, 2 and 4 pass through the $PBS_1$
and $PBS_2$, the total state of photons 1, 2, 3 and 4 evolves into
$$
|\Psi_1>=\alpha^2|H_{1^{\prime}}>|H_{3^{\prime}}>|H_{2^{\prime}}>|H_{4^{\prime}}>
$$
$$
+\beta^2|V_{1^{\prime}}>|V_{3^{\prime}}>|V_{2^{\prime}}>|V_{4^{\prime}}>
$$
$$
+\alpha\beta(|H_{1^{\prime}}>|V_{1^{\prime}}>|H_{2^{\prime}}>|V_{2^{\prime}}>
+|V_{3^{\prime}}>|H_{3^{\prime}}>|V_{4^{\prime}}>|H_{4^{\prime}}>)
\eqno{(4)}
$$
Alice rotate the polarization of photons $1^{\prime}$ by 45° with
four half-wave plate. The unitary transformation of photon through
the half wave plate is given by
$|H>\rightarrow\frac{1}{\sqrt{2}}(|H>+|V>)$ and
$|V>\rightarrow\frac{1}{\sqrt{2}}(|H>-|V>)$. After these
operations, the state(4) evolves into
$$
|\Psi_2>=\frac{\alpha^2}{\sqrt{2}}(|H_{1^{\prime}}>+|V_{1^{\prime}}>)|H_{3^{\prime}}>
|H_{2^{\prime}}>|H_{4^{\prime}}>
$$
$$
+\frac{\beta^2}{\sqrt{2}}(|H_{1^{\prime}}>-|V_{1^{\prime}}>)|V_{3^{\prime}}>|V_{2^{\prime}}>|V_{4^{\prime}}>
$$
$$
+\alpha\beta[\frac{1}{\sqrt{2}}(|2H_{1^{\prime}}>-|2V_{1^{\prime}}>)|H_{2^{\prime}}>|V_{2^{\prime}}>
+|V_{3^{\prime}}>|H_{3^{\prime}}>|V_{4^{\prime}}>|H_{4^{\prime}}>]
\eqno{(5)}
$$
To produce a maximally entangled four-photon state between Alice
and Bob, Photons $1^{\prime}$ and $4^{\prime}$ are incident
polarization beam splitter $PBS_3$ and $PBS_4$, the state become
$$
|\Psi_3>=\frac{\alpha^2}{\sqrt{2}}(|H_{5}>+|V_{6}>)|H_{3^{\prime}}>|H_{2^{\prime}}>|H_{7}>
$$
$$
+\frac{\beta^2}{\sqrt{2}}(|H_{5}>-|V_{6}>)|V_{3^{\prime}}>|V_{2^{\prime}}>|V_{8}>
$$
$$
+\alpha\beta[\frac{1}{\sqrt{2}}(|2H_{5}>-|2V_{6}>)|H_{2^{\prime}}>|V_{2^{\prime}}>
$$
$$
+|V_{3^{\prime}}>|H_{3^{\prime}}>)|H_{7}>|V_{8}>] \eqno{(6)}
$$
In order to delete the first two terms in Eq(6), Alice and Bob
pass the photons 5 and 6 through two symmetric beam splitter,
respectively. The second input ports of the beam splitters are
assumed to be single-photon states produced by single-photon
sources. After passing the symmetric beam splitters, the quantum
state of the auxiliary mode is measured. The outcome of this
BS-transformation is accepted only if the measurement of the
auxiliary mode gives the same number of photons like the ancilla
state was initially prepared: $1$ photon. The state of the system
is projected into
$$
|\Psi_4>=\frac{1}{2\sqrt{2}}(|2H_{5}>-|2V_{6}>)|H_{2^{\prime}}>|V_{2^{\prime}}>
$$
$$
+|V_{3^{\prime}}>|H_{3^{\prime}}>)|H_{7}>|V_{8}>] \eqno{(7)}
$$
Let photon modes 7 pass through  beam splitter, whose
transimmitance is $1/4$. The second input ports of the beam
splitters are assumed to be vacuum states. After passing the beam
splitters, the outcome of this BS-transformation is accepted only
if no photon is detected. we obtain
$$
|\Psi_5>=\frac{1}{\sqrt{2}}(|2H_{5}>-|2V_{6}>)|H_{2^{\prime}}>|V_{2^{\prime}}>
$$
$$
+|V_{3^{\prime}}>|H_{3^{\prime}}>)|H_{7}>|V_{8}>] \eqno{(8)}
$$
Let photon modes 5 and 6, 7 and 8 pass through polarization beam
splitter, we obtain
$$
|\Psi_5>=\frac{1}{\sqrt{2}}(|2H_{5^{\prime}}>-|2V_{5^{\prime}}>)|H_{2^{\prime}}>|V_{2^{\prime}}>
$$
$$
+|V_{3^{\prime}}>|H_{3^{\prime}}>)|V_{7^{\prime}}>|H_{7^{\prime}}>]
\eqno{(9)}
$$
Alice rotate the polarization of photons $5^{\prime}$ by 45° with
four half-wave plate. The state of the system is projected into
$$
|\Psi_6>=\frac{1}{4}(|H_{9}>|V_{9}>)|H_{2^{\prime}}>|V_{2^{\prime}}>
$$
$$
+\frac{1}{4}|V_{3^{\prime}}>|H_{3^{\prime}}>)|V_{7^{\prime}}>|H_{7^{\prime}}>]
\eqno{(10)}
$$
Let photon modes  $3^{\prime}$ and 9, $2^{\prime}$ and
$7^{\prime}$ pass through two polarization beam splitter,
respectively, we obtain
$$
|\Psi_7>=\frac{1}{\sqrt{2}}[|H_{10}>|V_{11}>)|H_{12}>|V_{13}>
+\frac{1}{4}|V_{10}>|H_{11}>)|V_{12}>|H_{13}>] \eqno{(11)}
$$
which is maximally entangled four-photon state. The probability of
success is $\frac{|\alpha\beta|^2}{8}$. Our scheme give the
generation of maximally entangled four-photon state in a
nondeterministic way. Note that maximally entangled four-photon
state have been obtained in \cite{pan,pp} from the two pairs of
entangled two-photon state, which is based on the four-photon
coincidence detection. Some quantum protocols have been designed
where entangled photon state are required before detection. Thus
post-selection cannot be
applied\cite{dik}. In this case, our scheme might find application.\\
To generate a maximally entangled two-photon state between Alice
and Bob, They rotate the polarization of the photon 11 and 13 by
45° with half wave plate HWP and obtain the state of the system
$$
|\Psi_8>=\frac{1}{2\sqrt{2}}(|H_{11}>|H_{13}>+|V_{11}>|V_{13}>)(|H_{10}>|H_{12}>+|V_{10}>|V_{12}>)
$$
$$
+\frac{1}{2\sqrt{2}}(|H_{11}>|V_{13}>+|V_{11}>|H_{13}>)(|H_{10}>|H_{12}>-|V_{10}>|V_{12}>)
\eqno{(12)}
$$
Let photon mode 11 and 13 pass through polarization beam
splitters, respectively. If Alice and Bob detect a single photon
at detector $D_4$ and $D_6$( or $D_5$ and $D_7$) and state of the
system is projected to
$$
|\Phi^{\dagger}>=\frac{1}{\sqrt{2}}(|H_{10}>|H_{12}>+|V_{10}>|V_{12}>)
\eqno{(13)}
$$
If they detect a single photon at detector $D_4$ and $D_7$( or
$D_5$ and $D_6$) and state of the system is projected to
$$
|\Phi^{-}>=\frac{1}{\sqrt{2}}(|H_{10}>|H_{12}>-|V_{10}>|V_{12}>)
\eqno{(14)}
$$
In this case, they can easily transform it into the form of
$|\Phi^{\dagger}>$.\\ In what follows, We show the scheme also can
be used to purify mixed polarization-entangled state of the form
$$\gamma|\Psi^{\dagger}><\Phi^{\dagger}|+(1-\gamma)|VV><VV|
\eqno{(15)}
$$ here
$|\Psi^{\dagger}>=\frac{1}{\sqrt{2}}(|H,V>+|V,H>)$ and
$\gamma>\frac{1}{2}$. Let us assume that Alice and Bob are given
two pairs of photons in the following polarization entangled state
$$
\rho_{12}=\gamma|\Psi^{\dagger}>_{12}{_{12}<}\Phi^{\dagger}|+(1-\gamma)|V_1>|V_2><V_1|<V_2|
\eqno{(16)}
$$
$$
\rho_{34}=\gamma|\Psi^{\dagger}>_{34}{_{34}}<\Phi^{\dagger}|+(1-\gamma)|V_3>|V_4><V_3|<V_4|
\eqno{(17)}
$$
Alice holds photons 1 and 3, and Bob holds photon 2 and 4 and let
photon mode 1, 2, 3 and 4 pass through four beam splitters,
respectively and spatially separate the vertical and horizontal
polarization modes of the four separate modes. The state of the
system evolve into
$$
\rho=\gamma^2|\Psi^{\dagger}>_{5678}
|\Psi^{\dagger}>_{5^{\prime}6^{\prime}7^{\prime}8^{\prime}}
{_{5678}}<\Phi^{\dagger}|{_{5^{\prime}6^{\prime}7^{\prime}8^{\prime}}}<\Phi^{\dagger}|
$$
$$
+(1-\gamma)^2|V_6>|V_8>|V_{6^{\prime}}>|V_{8^{\prime}}>
<V_6|<V_8|<V_{6^{\prime}}|<V_{8^{\prime}}|
$$
$$
+\gamma(1-\gamma)|\Psi^{\dagger}>_{5678}|V_{6^{\prime}}>|V_{8^{\prime}}>
{_{5678}}<\Phi^{\dagger}|<V_{6^{\prime}}|<V_{8^{\prime}}|
$$
$$
+\gamma(1-\gamma)|\Psi^{\dagger}>_{5^{\prime}6^{\prime}7^{\prime}8^{\prime}}
|V_6>|V_8>{_{5^{\prime}6^{\prime}7^{\prime}8^{\prime}}}<\Phi^{\dagger}|<V_6|<V_8|
\eqno{(18)}
$$
here
$$
|\Psi^{\dagger}>_{5678}=\frac{1}{\sqrt{2}}(|H_5>|V_8>+|V_6>|H_7>)
$$
$$
|\Psi^{\dagger}>_{{5^{\prime}6^{\prime}7^{\prime}8^{\prime}}}=\frac{1}{\sqrt{2}}(|H_{5^{\prime}}>|V_{8^{\prime}}>+|V_{6^{\prime}}>|H_{7^{\prime}}>)
$$
Let photon modes $5$ and $5^{\prime}$, $6$ and $6^{\prime}$, $7$
and $7^{\prime}$, $8$ and $8^{\prime}$ pass through four symmetric
beam splitters, respectively and the eight output modes of four
symmetric beam splitters are then incident eight symmetric beam
splitters. The second input ports of the eight beam splitters are
assumed to be single-photon states produced by single-photon
sources. After passing the symmetric beam splitters, the quantum
state of the auxiliary mode is measured. The outcome  is accepted
only if the measurement of the auxiliary mode gives the single
photon state. The (unnormalized) state of the system is projected
into
$$
|\Phi_1>=\gamma^2(|\psi_1> |\psi_2>+|\psi_3>
|\psi_4>)(<\phi_1|<\phi_2|+<\phi_3|<\phi_4|)
$$
$$
+(1-\gamma)^2|\psi_2> |\psi_3><\phi_2|<\phi_3| \eqno{(19)}
$$
where
$$
|\psi_1>=\frac{1}{\sqrt{2}}(|2H_{10}>-|2H_{11}>)
$$
$$
|\psi_2>=\frac{1}{\sqrt{2}}(|2V_{16}>-|2V_{17}>)
$$
$$
|\psi_3>=\frac{1}{\sqrt{2}}(|2V_{12}>-|2V_{13}>)
$$
$$
|\psi_4>=\frac{1}{\sqrt{2}}(|2H_{14}>-|2H_{15}>)
$$
Let photon modes $10$ and $11$, $12$ and $13$, $14$ and $15$, $16$
and $17$ pass through the four symmetric beam splitter, the state
of the system become
$$
|\Phi_2>=\gamma^2(|H_{10^{\prime}}>|H_{11^{\prime}}>|V_{16^{\prime}}>|V_{17^{\prime}}>
+|V_{12^{\prime}}>|V_{13^{\prime}}>|H_{14^{\prime}}>|H_{15^{\prime}}>)
$$
$$
\times
(<H_{10^{\prime}}|<H_{11^{\prime}}|<V_{16^{\prime}}|<V_{17^{\prime}}|+
<V_{12^{\prime}}|<V_{13^{\prime}}|<H_{14^{\prime}}|<H_{15^{\prime}}|)
$$
$$
+(1-\gamma)^2|V_{12^{\prime}}>|V_{13^{\prime}}>|V_{16^{\prime}}>|V_{17^{\prime}}>
<V_{12^{\prime}}|<V_{13^{\prime}}|<V_{16^{\prime}}|<V_{17^{\prime}}|
\eqno{(20)}
$$
Let photons modes $10^{\prime}$ and $12^{\prime}$, $11^{\prime}$
and $13^{\prime}$, $14^{\prime}$ and $16^{\prime}$, $15^{\prime}$
and $17^{\prime}$ incident four polarization beam splitter,
respectively. We obtain
$$
|\Phi_3>=\gamma^2(|H_{a}>|H_{b}>|V_{c}>|V_{d}>
+|V_{a}>|V_{b}>|H_{c}>|H_{d}>)
$$
$$
\times
 (<H_{a}|<H_{b}|<V_{c}|<V_{d}|+
<V_{a}|<V_{b}|<H_{c}|<H_{d}|)
$$
$$
+(1-\gamma)^2|V_{a}>|V_{b}>|V_{c}>|V_{d}>
<V_{a}|<V_{b}|<V_{c}|<V_{d}|\eqno{(21)}$$. Now Alice and Bob
rotate the polarization of the photon mode b and d by 45° with
half wave plate and let output mode of half-wave plate incident on
the two polarization beam splitter, respectively. If Alice and Bob
detect a single photon at detector $D_9$ and $D_{11}$ ( or
$D_{10}$ and $D_{12}$) and state of the system is projected to
$$
\rho=\gamma^{\prime}|\Psi^{\dagger}><\Phi^{\dagger}|+(1-\gamma^{\prime})|VV><VV|
\eqno{(22)}
$$
with a large fraction
$\gamma^{\prime}=\frac{\gamma^2}{(1-\gamma)^2+\gamma^2}>\gamma$ of
the entangle two-photon state in $|\Psi^{\dagger}>$. If they
detect a single photon at detector $D_9$ and $D_{12}$( or $D_{10}$
and $D_{11}$) and state of the system is projected to
$$
\rho=\gamma^2|\Psi^{-}><\Phi^{-}|+(1-\gamma)^2|VV><VV| \eqno{(23)}
$$
In this case, they can easily transform it into the form of
Eq.(22). Thus as long as the initial ensemble is sufficiently
large, iterating the procedure will yield a entangled two-photon
state arbitrary close to entangled state $|\Psi^{\dagger}>$. Note
that our scheme  can also be used to purify the mixed entangled
state of the form
$\gamma^2|\Psi^{\dagger}><\Phi^{\dagger}|+(1-\gamma)^2|\Phi^{\dagger}><\Phi^{\dagger}|$
, here $|\Phi>_{12}=\frac{1}{\sqrt{2}}(|HH>+|VV>)$. If
$\gamma>1/2$, the state can be purified to $|\Psi^{\dagger}>$, For
$\gamma|1/2$, the state can be purified to $|\Phi^{\dagger}>$.
This conclude our discussion on purifying mixed entangled
two-photon state.\\
In conclusion, we have presented a concentration and purification
scheme for nonmaximally entangled pure  and mixed porization
entangled state. We firstly showed that two distant parties Alice
and Bob first start with two shared but less entangled photon pure
states to produce a four photon GHZ state in a nondeterministic
way, and then perform a $45°$ polarization measurement onto one of
the two photons at each location such that the remaining two
photon are projected onto a maximally entangled state. We further
show the scheme also can be used to purify a kind of mixed
polarization-entangled state. Note that maximally entangled
four-photon state have been obtained in \cite{pan,pp} from the two
pairs of entangled two-photon state, which is based on the
four-photon coincidence detection. Some quantum protocols have
been designed where entangled photon state are required before
detection\cite{dik}. Thus post-selection cannot be applied. In
this case, our scheme might find application. One of the main
difficulties of this scheme in respect to an experimental
demonstration is the availability of photon-number sources.
Another difficulty consists in the requirement on the sensitivity
of the detectors. These detectors should be capable of
distinguishing between no photon, one photon or more photons.

\begin{flushleft}

{\Large \bf Figure Captions}

\vspace{\baselineskip}

{\bf Figure 1.} The schematic diagram of our scheme for
entanglement concentration. $PBS_i$ denote polarization beam
splitters and $HWP$ denote half-wave plate. $BS$ is beam splitter
and $D_i$ are photon number detectors.

{\bf Figure 2.} The schematic is shown to purifying mixed
polarization entangled two-photon state. $PBS$ denote polarization
beam splitters and $HWP$ denote half-wave plate. $BS$ is beam
splitter and $D_i$ are photon number detectors.
\end{flushleft}


\begin{thebibliography}{99}
\bibitem{chg}C.H. Bennett, G. Brassard, C. Crépeau, R. Jozsa, A. Peres, and W.K. Wootters, Phys. Rev. Lett. 70, 1895 (1993).
\bibitem{ake}A.K. Ekert, Phys. Rev. Lett. 67, 661 (1991).
\bibitem{chs}C.H. Bennett and S.J. Wiesner, Phys. Rev. Lett. 69, 2881 (1992).
\bibitem{chd}D.P. Divincenzo, Science 270, 255 (1995); C.H. Bennett and D.P. Divincenzo, Nature (London) 404, 247 (2000)
\bibitem{con}C.H. Bennett H.J. Bernstein, S. Popescu, and B. Schumacher, Phys. Rev. A 53, 2046 (1996);
C.H. Bennett G. Brassard, S. Popescu, B. Schumacher, J.A. Smolin,
and W.K. Wootters, Phys. Rev. Lett. 76, 722 (1996); D. Deutsch,
A.K. Ekert, R. Jozsa, C. Macchiavello, S. Popescu, and A. Sanpera,
Phys. Rev. Lett. 77, 2818 (1996); M. Murao, M.B. Plenio, S.
Popescu, V. Vedral, and P.L. Knight, Phys. Rev. A 57, R4075
(1998).
\bibitem{pan}Zhi Zhao, Jian-Wei Pan and M. S. Zhan, Phys. Rev. A 64, 014301 (2001)
\bibitem{aa}
Takashi Yamamoto, Masato Koashi, and Nobuyuki Imoto, Phys. Rev. A
64, 012304 (2001)
\bibitem{pa}J.-W. Pan, C. Simon, C. Brukner, and A. Zeilinger, Nature (London) 410, 1067 (2001).
\bibitem{zou} XuBo Zou, K. Pahlke, W. Mathis, quant-ph/0110168
\bibitem{pp}D. Bouwmeester, J.-W. Pan, M. Daniell, H. Weinfurter, and A. Zeilinger, Phys. Rev. Lett. 82, 1345 (1999).
Jian-Wei Pan, Matthew Daniell, Sara Gasparoni, Gregor Weihs, and
Anton Zeilinger , Phys. Rev. Lett. 86, 4435 (2001)
\bibitem{dik}Dik Bouwmeester, Phys. Rev. A 63, 040301 (2001)
\end{thebibliography}
\end{document}